# Bayesian Network Enhanced with Structural Reliability Methods: Methodology

Daniel Straub[1] & Armen Der Kiureghian[2]

## Abstract

We combine Bayesian networks (BNs) and structural reliability methods (SRMs) to create a new computational framework, termed enhanced Bayesian network (eBN), for reliability and risk analysis of engineering structures and infrastructure. BNs are efficient in representing and evaluating complex probabilistic dependence structures, as present in infrastructure and structural systems, and they facilitate Bayesian updating of the model when new information becomes available. On the other hand, SRMs enable accurate assessment of probabilities of rare events represented by computationally demanding, physically-based models. By combining the two methods, the eBN framework provides a unified and powerful tool for efficiently computing probabilities of rare events in complex structural and infrastructure systems in which information evolves in time. Strategies for modeling and efficiently analyzing the eBN are described by way of several conceptual examples. The companion paper applies the eBN methodology to example structural and infrastructure systems.

---

[1] Associate Professor, Engineering Risk Analysis Group, Technische Universität München, Arcisstr. 21, 80290 München, Germany. Email: straub@tum.de

[2] Taisei Professor of Civil Engineering, Dept. of Civil & Environmental Engineering, Univ. of California, Berkeley, CA 94720. Email: adk@ce.berkeley.edu

# 1 Introduction

Structural reliability methods (SRMs) have been developed and successfully applied in the engineering community to solve for the probability of an event $E$ that is given through an integral of the form

$$\Pr(E) = \int_{\mathbf{x} \in \Omega_E(\mathbf{x})} f(\mathbf{x}) d\mathbf{x} \tag{1}$$

The event $E$ is defined as a domain $\Omega_E$ in the outcome space of random variables $\mathbf{X} = (X_1, X_2, \ldots, X_n)$, which are specified through their joint probability density function (PDF) $f(\mathbf{x})$. In structural system reliability theory (Ditlevsen and Madsen, 1996), $\Omega_E$ is defined in terms of a set of continuous and continuously differentiable limit-state functions $g_i(\mathbf{x})$, $i = 1, \ldots, m$, in the form

$$\Omega_E(\mathbf{x}) = \left\{ \min_{1 \leq k \leq K} \left[ \max_{i \in C_1} g_i(\mathbf{x}), \ldots, \max_{i \in C_K} g_i(\mathbf{x}) \right] \leq 0 \right\} \tag{2}$$

where $C_k$ is an index set denoting the $k$-th cut set of the system. The problem is said to be a "component" reliability problem when $m = 1$; a "parallel-system" reliability problem when $K = 1$; and a "series-system" reliability problem when each cut set contains only one index. The form in (1) corresponds to a "general system" reliability problem.

Occasionally, the structural reliability problem is defined in the total-probability form

$$\Pr(E) = \int_{\mathbf{x}} p_E(\mathbf{x}) f(\mathbf{x}) d\mathbf{x} \tag{3}$$

where $p_E(\mathbf{x})$ is the conditional probability of event $E$ given $\mathbf{X} = \mathbf{x}$. Provided $p_E(\mathbf{x})$ is continuously differentiable with respect to $\mathbf{x}$, one can easily show (Wen and Chen, 1987) that the above form reduces to Equation (1) with $\Omega_E(\mathbf{x}) = \{x_\varphi + \beta(\mathbf{x}) \leq 0\}$, where $x_\varphi$ is the outcome of a standard normal random variable $X_\varphi$. $\beta(\mathbf{x}) = \Phi^{-1}[1 - p_E(\mathbf{x})]$ is the conditional reliability index given $\mathbf{X} = \mathbf{x}$, wherein $\Phi^{-1}[\cdot]$ denotes the inverse of the standard normal cumulative probability function. This form is applicable when random variables exogenous to $\mathbf{X}$ also influence event $E$.

Methods for solving integrals of the form in (1) include the First- and Second-Order Reliability Methods (FORM and SORM) and a variety of simulation approaches, including crude Monte Carlo, importance sampling, directional sampling and subset simulation. In the following we will refer to this class of methods as SRMs. These methods are well-documented in a variety of textbooks and articles, including (Ditlevsen and Madsen, 1996), (Rackwitz, 2001) and (Der Kiureghian, 2005), and are implemented in a number of educational and commercial software (Ellingwood, 2006).

Bayesian networks (BNs), also known as belief networks, are probabilistic models that facilitate efficient representation of the dependence structure among random variables by graphical means. BNs have been developed during the past 25 years, mostly in the field of artificial intelligence, for representing probabilistic information and reasoning (Russell and Norvig, 2003). They have found applications in many fields such as statistical modeling, language processing, image recognition and machine learning, and have begun to be used in engineering risk analysis. Recent applications in this field are reported, e.g., in (Friis-Hansen, 2000; Faber *et al.*, 2002; Friis-Hansen, 2004; Mahadevan and Rebba, 2005; Grêt-Regamey and Straub, 2006; Nishijima *et al.*, 2009; Bensi *et al.*, 2009). Since the BN methodology is likely to be new to most readers of this journal, its essential elements are briefly introduced in a separate section.

As would be expected, both SRMs and BNs have advantages and limitations. SRMs are applicable to continuous random variables with a known joint distribution. Any form of statistical dependence can be handled. However, several SRMs, e.g., FORM and SORM, are not suitable for discrete random variables. Furthermore, these methods are difficult to apply for non-experts, particularly in the context of information updating, and are not easily presentable in a graphical form. They cannot generally be included in automated algorithms that can run without an expert. On the other hand, the BN is highly effective for analyzing discrete random variables and for information updating. It is also an effective tool for decision-making, and its graphical form provides a concise representation of statistical dependence that can be understood also by non-experts. In its discrete form, the BN can be run by the lay engineer, even in an automated mode for near-real time decision support. However, for continuous random variables, the BN has practical limitations on the type of distributions and the form of statistical dependence that can be handled. Furthermore, it is not ideally suited for computing small probabilities, which is the specialty of SRMs. It follows that a combination of the two methods can potentially be a powerful tool for probabilistic analysis and decision-making.

So far, few publications have considered the use of BNs in civil engineering risk analysis from a methodological viewpoint. Friis-Hansen (Friis-Hansen, 2000) combines example applications with a discussion of some of the methodological issues involved, such as the discretization of random variables. A SRM is used to compute the conditional probability tables of the BN, similar to what is proposed in this paper, yet without a formal framework. In (Friis-Hansen, 2004), an application of BN to structural reliability problems is investigated, based on discretizing all random variables. Additionally, a number of authors discuss the modeling of reliability problems using BNs (Bobbio *et al.*, 2001; Mahadevan and Rebba, 2005; Langseth and Portinale, 2007; Neil *et al.*, 2008), but these do not include structural reliability applications.

The objective of this paper is to explore the possibility of combining the SRM and BN into an enhanced tool for probabilistic analysis and decision-making. Denoted enhanced Bayesian Network or eBN, the proposed tool is a BN which has both discrete and continuous nodes with arbitrary distributions and interdependencies. We explore the rules and computations that are necessary to reduce an eBN into a reduced BN (rBN) with discrete nodes only, for which existing exact methods of inference can be used. The reduction is performed through a process of elimination of continuous nodes. We show that the required computations can be performed by an SRM through background analyses, which can remain hidden to the user of the rBN. Various alternatives for eliminating continuous nodes so as to optimally produce the rBN are explored. Finally, modeling strategies that enable efficient computations are presented and the limitations of the approach are discussed. The application of the methodology is shown in a companion paper, considering reliability analysis of an individual structural system and a system of structural systems.

## 2 Bayesian networks

We limit ourselves to introducing the most important concepts of the BN as required for the remainder of the paper. For an extensive introduction to BN, we refer to the standard textbook of Jensen and Nielsen (Jensen and Nielsen, 2007) and the review paper on the application of BN for engineering reliability applications by Langseth and Portinale (Langseth and Portinale, 2007). Broader introductions to BN and related concepts of probabilistic knowledge representation and reasoning are provided by (Pearl, 1988) and (Russell and Norvig, 2003).

BNs are probabilistic models based on directed acyclic graphs. They represent a probability measure $\pi(\mathbf{z})$ over the outcome space of a set of random variables $\mathbf{Z} = (Z_1,...,Z_N)$. Each variable $Z_i$ can be defined in a discrete and finite outcome space (discrete random variable) or a continuous outcome space (continuous random variable). For discrete random variables, $\pi(\mathbf{z}) = \Pr(\mathbf{Z} = \mathbf{z}) = p(\mathbf{z})$ is the joint probability mass function (PMF). For continuous random variables, $\pi(\mathbf{z}) = \partial^N \Pr(\mathbf{Z} \leq \mathbf{z})/\partial \mathbf{z} = f(\mathbf{z})$ is the joint PDF. The size of the joint outcome space of $\mathbf{Z}$ for which $\pi(\mathbf{z})$ must be defined increases exponentially with the number of variables, but the BN enables an efficient modeling by factoring the joint probability distribution into conditional (local) distributions for each variable given its *parents*. A simple BN with five variables is illustrated in Figure 1, where $Z_1$ is a parent of $Z_3$ and $Z_4$, and $Z_2$ is a parent of $Z_4$ and $Z_5$.

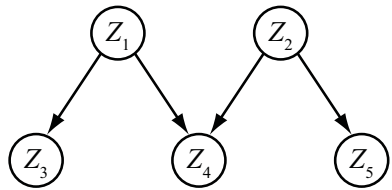

*Figure 1. A simple Bayesian network.*

The joint probability measure for this network is given as

$$\pi(\mathbf{z}) = \pi(z_1, z_2, z_3, z_4, z_5) = \pi(z_1)\pi(z_2)\pi(z_3|z_1)\pi(z_4|z_1,z_2)\pi(z_5|z_2) \qquad (4)$$

which can be written in the compact and general form

$$\pi(\mathbf{z}) = \prod_{i=1}^{N} \pi\left[z_i \mid pa(Z_i)\right] \tag{5}$$

where $pa(Z_i)$ denotes the set of parents of $Z_i$. In addition, let $ch(Z_i)$ denote the *children* of $Z_i$ and $sp(Z_i)$ the *spouses* of $Z_i$. The spouses are all variables that share a child with $Z_i$ but are not children of $Z_i$. For example, in the BN shown in Figure 1, $pa(Z_1)$ is the empty set, $ch(Z_1) = \{Z_3, Z_4\}$ and $sp(Z_1) = \{Z_2\}$. The parents, children and spouses of a variable together form the *Markov blanket* of that variable,

$$bl(Z_i) = pa(Z_i) \cup ch(Z_i) \cup sp(Z_i) \tag{6}$$

The Markov blanket is an important concept, since, for given values of $bl(Z_i)$, $Z_i$ is statistically independent of all other variables, i.e.,

$$\pi\left(z_i \mid \mathbf{z}_{-i}\right) = \pi\left[z_i \mid bl(Z_i)\right] \tag{7}$$

where $\mathbf{z}_{-i}$ denotes realizations of all variables other than $Z_i$. This independence relation follows directly from the *d-separation* rules formulated by Pearl (Pearl, 1988), which represent the independence assumptions encoded in the graphical structure of the BN. When evidence (information) is available on a set of variables in the BN, the distributions of all remaining variables are updated using Bayes' rule. The rules of d-separation help to identify those variables whose probability measure change upon the evidence. As an example, if the evidence $\{Z_5 = e_5\}$ is available on the BN in Figure 1, the marginal probability measures on the variables $Z_2$ and $Z_4$ will change, whereas those of variables $Z_1$ and $Z_3$ will not change because they are d-separated from $Z_5$. As an example, $\pi(z_4 | e_5)$ is computed as

$$\pi(z_4|e_5) = \frac{\pi(z_4, e_5)}{\pi(e_5)}$$

$$= \frac{\sum_{Z_1}\sum_{Z_2} \pi(z_4|z_1, z_2)\pi(e_5|z_2)\pi(z_1)\pi(z_2)}{\sum_{Z_2} \pi(e_5|z_2)\pi(z_2)} \quad (8)$$

$$= \frac{\sum_{Z_2} \pi(e_5|z_2)\pi(z_2)\sum_{Z_1} \pi(z_4|z_1, z_2)\pi(z_1)}{\sum_{Z_2} \pi(e_5|z_2)\pi(z_2)}$$

Equation (8) is for the case of discrete random variables; for continuous random variables the summation operations must be replaced with integration operations. From the numerator in the last line of Equation (8), we can deduce the principle behind exact inference algorithms, which are available for solving BNs with discrete random variables (Lauritzen and Spiegelhalter, 1988; Shenoy and Shafer, 1990; Zhang and Poole, 1996; Dechter, 1996; Jensen and Nielsen, 2007). In essence, these algorithms aim to find an optimal ordering of the summation operations requiring the lowest CPU time and/or storage capacity. Available algorithms perform differently depending on the application. It has been shown that finding an optimal ordering is an NP-complete problem (Cooper, 1990). This concept from computer science indicates that it is not promising to search for an algorithm that finds the optimal ordering in all cases. Several algorithms have been implemented in a number of commercially or freely available software (see Murphy Murphy, 2001 for an overview). Detailed knowledge of these algorithms is not necessary for this paper.

To appreciate the difficulty of the problem, observe that the computation in Equation (8) requires multiplication operations in the joint outcome space of $Z_1$, $Z_2$ and $Z_4$. In the language of (Jensen and Nielsen, 2007), the variables $Z_1$, $Z_2$ and $Z_4$ form the largest *clique* for this BN, thus necessitating the computation of the *potential* in the domain of these three variables, potentials being tables of conditional probabilities. Since the size of the potential increases exponentially with the number of variables involved, the size of the largest potential to handle is critical for the performance of the algorithm. Consider a network consisting of one child node with 20 parent nodes. General inference will require manipulating probabilities in the joint space of all these variables (i.e., the largest clique involves all 21 variables). Even if each variable has only two states, the potential to handle already contains $2^{21}$ entries!

Exact inference algorithms exist also for two special cases of so-called hybrid BNs, which involve both continuous and discrete random variables. The first case is BNs with nodes that are defined as Gaussian random variables, whose means are linear functions of their parents. The application of such BNs is rather limited, in particular since the continuous nodes must not have any discrete children. The second case is BNs whose nodes are defined as mixtures of truncated exponentials (MTE). Such MTEs can be thought of as an extension of discrete random variables, whereby the probability density within each interval is approximated by a linear combination of exponential functions instead of a constant (Langseth *et al.*, 2009).

As an alternative to exact inference, approximate inference algorithms using simulation techniques have been developed, of which Markov Chain Monte Carlo (MCMC) methods (Gilks *et al.,* 1996; Beck and Au*,* 2002) have become especially popular. Combinations of exact and approximate inference algorithms are also available (see, e.g., Murphy*,* 2002). We will briefly return to discuss approximate inference, but otherwise restrict this paper to exact inference in BNs with discrete random variables.

## 3  Framework for an enhanced Bayesian Network

### 3.1  Definitions

We define as enhanced Bayesian networks (eBNs) a subclass of BNs that have the following properties:

a) The BN has nodes that are defined in a finite sample space (discrete nodes) and nodes that represent vectors of continuous random variables (continuous nodes).

b) The states of each discrete node that is a child of at least one continuous node are defined as domains in the outcome space of its parents, in which case the node is deterministic, or are defined by a PMF that is parameterized by the parent nodes, in which case the node is random.

More formally, let us define a set of discrete random variables $\mathbf{Y} = \{Y_1, \ldots, Y_{n_Y}\}$ and a set of vectors of (not necessarily independent) continuous random variables $\mathbf{X} = \{\mathbf{X}_1, \ldots, \mathbf{X}_{n_X}\}$. The complete set of nodes in the network is denoted $\mathbf{Z} = \{\mathbf{X}, \mathbf{Y}\}$. A discrete variable $Y_i$ has a set of possible states $y_i^{(k_i)}$, where $i$ is the index for the variable and $k_i$ is the index of the state, and a conditional PMF $p[y_i^{(k_i)} | pa(Y_i)] = \Pr[Y_i = y_i^{(k_i)} | pa(Y_i)]$. To enhance readability, we omit the state index $k_i$ whenever there is no ambiguity. A continuous node with index $i$ represents a vector of continuous random variables $\mathbf{X}_i = (X_{i,1}, \ldots, X_{i,n_i})$ and is characterized by the joint conditional PDF $f[\mathbf{x}_i | pa(\mathbf{X}_i)]$. The BN is then characterized by the combined measure

$$p(\mathbf{y}|\mathbf{x}) f(\mathbf{x}) = \prod_{Y_i \in \mathbf{Y}} p\left[y_i \big| pa(Y_i)\right] \prod_{\mathbf{X}_i \in \mathbf{X}} f\left[\mathbf{x}_i \big| pa(\mathbf{X}_i)\right] \tag{9}$$

When the discrete node $Y_i$ has exclusively continuous parents with realizations denoted by $\mathbf{x}_{pa(Y_i)}$, the conditional PMF $p[y_i | pa(Y_i)]$ is defined according to Equations (1) or (3). Thus, each state $y_i^{(k_i)}$ is defined through a domain $\Omega_i^{(k_i)}(\mathbf{x}_{pa(Y_i)})$. If the node $Y_i$ additionally has discrete parents $\mathbf{y}_{pa(Y_i)}$, this function must be defined for each combination of the states of $\mathbf{y}_{pa(Y_i)}$ separately and the corresponding notation of the domain is $\Omega_{i,k_{pa}}^{(k_i)}(\mathbf{x}_{pa(Y_i)})$, wherein $k_{pa}$ denotes the joint state of the discrete parents of $Y_i$. As an example, $\mathbf{y}_{pa(Y_i)}$ may describe the operational modes of a mechanical system and $\Omega_{i,k_{pa}}^{(k_i)}(\mathbf{x}_{pa(Y_i)})$ represent the models describing the system performance for different modes $k_{pa}$; or $\mathbf{y}_{pa(Y_i)}$ may be a variable representing whether or not a wave hits the deck of an offshore platform, and $\Omega_{i,k_{pa}}^{(k_i)}(\mathbf{x}_{pa(Y_i)})$ represent the appropriate structural models for each case.

All discrete nodes in the eBN are defined as single variables. This is no limitation, since for discrete variables it is straightforward to transform the joint space of several variables into the space of a single variable. As an example, if variable $Y_1$ has two states 0 and 1, and variable $Y_2$ has two states 0 and 1, the joint space of the two variables can be represented by a single variable with four discrete states 00, 01, 10 and 11.

## 3.2 Inference problem and solution strategy

The inference problem considered here is that of determining $p(\mathbf{y}_j | \mathbf{y}_e)$, where $\mathbf{Y}_j$ are the random variables of interest and $\mathbf{Y}_e = \mathbf{y}_e$ is the available evidence. $\mathbf{Y}_j$ and $\mathbf{Y}_e$ are both subsets of $\mathbf{Y}$, the set of discrete random variables. This is a restriction of the general case, in that all variables of interest are discrete random variables and all evidence is on discrete random variables. However, for many applications these restrictions are not critical. As we will show, equivalent discrete variables can be introduced for any continuous variable of interest. Furthermore, as described later, certain types of evidence on continuous random variables can be handled by introducing a binary discrete random variable with one of its states corresponding to the observed evidence.

We propose to solve inference problems of the type described above through a two-step procedure. The first step is the determination of $p(\mathbf{y})$, the joint PMF of the discrete variables $\mathbf{Y}$. $p(\mathbf{y})$ is obtained through elimination of the continuous nodes in the eBN and is represented by a BN itself. The resulting reduced BN is referred to as rBN. In the second step, since the rBN consists only of discrete nodes, $p(\mathbf{y}_j | \mathbf{y}_e)$ is evaluated from $p(\mathbf{y})$ by use of existing algorithms for exact inference.

The motivation for transforming the eBN into a rBN consisting only of discrete nodes is the availability of exact and easy-to-use inference algorithms for BNs with discrete nodes, and the possibility to utilize well established SRMs. An alternative strategy that is not investigated here is the use of approximate algorithms that can perform inference directly in hybrid BNs, i.e., BNs with both continuous and discrete nodes (Langseth *et al.*, 2009). In the companion paper (Straub and Der Kiureghian*, 2010*) we use simulation-based algorithms for hybrid BNs to check our models, but it is noted that such methods are either computationally inefficient in the general case (e.g., in the case of rejection or likelihood sampling), or they have unknown rates of convergence (e.g., in the case of MCMC), making them difficult to employ in automated algorithms. By automated algorithms we mean algorithms that can be included in software and then be efficiently applied by engineers and other experts who are not specialists in the underlying probabilistic modeling and algorithms. Automated algorithms are particularly relevant for near-real-time decision support systems. The principle advantage of the proposed computation strategy is that only the development of the rBN is time-consuming and requires specialist knowledge in probability and reliability analysis. By implementing the resulting rBN in software that allows exact inference, the resulting model can be applied by non-specialists. Furthermore, the rBN is easily extended to a decision graph, thus allowing direct decision optimization. We believe that the development of such automated algorithms will advance the dissemination of probabilistic methods in practice for a variety of complex engineering decision problems under uncertainty.

## 3.3 Determining the rBN

In order to establish the rBN, the continuous nodes $\mathbf{X}_i$ are removed from the network. An algorithm for elimination of nodes in an influence diagram, which includes the BN as a special case, is described in (Shachter, 1988, Shachter, 1986).

### 3.3.1 Node elimination algorithm

Following (Shachter, 1986), we first define as barren nodes all random variables without children that do not receive any evidence. If $\mathbf{X}_i$ is a barren node in an eBN, we can simply remove it together with the links directing to it, without changing $p(\mathbf{y})$.

Second, consider theorem 2 from (Shachter, 1986) that describes the conditions for reversing a directed link (arc): "*Given that there is an arc (i,j) between chance nodes i and j, but no other directed (i,j)-path in a regular influence diagram, arc (i,j) can be replaced by arc (j,i). Afterward, both nodes inherit each other's conditional predecessors.*" According to this theorem, a directed link between two nodes (from node *i* to node *j*) can be reversed if there is no other path in the same direction (otherwise the new network would become cyclic), by adding directed links from all parents of node *i* to node *j* and from all parents of node *j* to node *i*.

The process of eliminating node $\mathbf{X}_i$ in the eBN proceeds by first reversing all directed links from $\mathbf{X}_i$ to $ch(\mathbf{X}_i)$, the children of $\mathbf{X}_i$, until $ch(\mathbf{X}_i)$ is the empty set and $\mathbf{X}_i$ is a barren node. Then, $\mathbf{X}_i$, together with all the links pointing to it, can simply be removed. Figure 2 shows an example of the process. The order of the reversing operations can be chosen freely as long as it is ensured that the resulting network is acyclic at any stage; in Figure 2, the link from $\mathbf{X}_1$ to $Y_6$ cannot be reversed first, as this would lead to the cycle $Y_5 \rightarrow Y_6 \rightarrow \mathbf{X}_1 \rightarrow Y_5$. Later, we will show that the order of the reversing operations can influence the form of the rBN, and we will address the optimal ordering. It is noted that upon elimination of the continuous nodes, discrete nodes that were defined as deterministic functions of their parents (through domains $\Omega_i$) become random nodes, encapsulating the uncertainty in their continuous parent nodes.

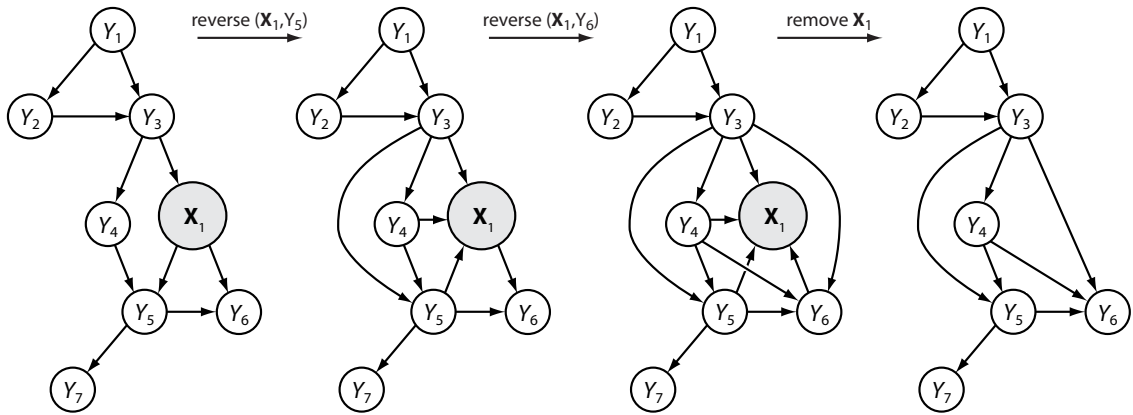

*Figure 2. Illustration of an enhanced Bayesian network and a link reversal sequence for removal of node $\mathbf{X}_1$ to arrive at the rBN.*

### 3.3.2 Computing the conditional probability tables (potentials) of the rBN

For the given structure of the rBN, as determined through the node elimination algorithm, it is necessary to compute the conditional probability tables (potentials) of the variables. Here we show that these computations can be performed through a SRM. To distinguish between the eBN and the rBN structure, let $pa(Y_i)$ denote the parents of variable $Y_i$ in the eBN and let $pa'(Y_i)$ denote its parents in the rBN.

Let $\mathbf{Y}_C$ denote all discrete variables that are children of at least one continuous variable in the eBN, $\mathbf{Y}_C = \{\mathbf{Y} \cap [ch(\mathbf{X}_1) \cup \ldots \cup ch(\mathbf{X}_{n_\mathbf{X}})]\}$, and let $\mathbf{Y}_{NC}$ denote all remaining discrete variables. It follows from the node elimination algorithm that all variables in $\mathbf{Y}_{NC}$ will have the same parents in the rBN as in the original eBN, and therefore the same conditional probability tables. This is also evident when formulating the joint PMF of all discrete nodes for the general case:

$$\begin{aligned} p(\mathbf{y}) &= \int_{\mathbf{X}_1} \ldots \int_{\mathbf{X}_{n_\mathbf{X}}} p(\mathbf{y} \mid \mathbf{x}) f(\mathbf{x}) d\mathbf{x}_1 \ldots d\mathbf{x}_{n_\mathbf{X}} \\ &= \int_{\mathbf{X}_1} \ldots \int_{\mathbf{X}_{n_\mathbf{X}}} \prod_{Y_i \in \mathbf{Y}} p\left[y_i \mid pa(Y_i)\right] \prod_{X_i \in \mathbf{X}} f\left[\mathbf{x}_i \mid pa(\mathbf{X}_i)\right] d\mathbf{x}_1 \ldots d\mathbf{x}_{n_\mathbf{X}} \\ &= \prod_{Y_i \in \mathbf{Y}_{NC}} p\left[y_i \mid pa(Y_i)\right] \int_{\mathbf{X}_1} \ldots \int_{\mathbf{X}_{n_\mathbf{X}}} \prod_{Y_i \in \mathbf{Y}_C} p\left[y_i \mid pa(Y_i)\right] \prod_{X_i \in \mathbf{X}} f\left[\mathbf{x}_i \mid pa(\mathbf{X}_i)\right] d\mathbf{x}_1 \ldots d\mathbf{x}_{n_\mathbf{X}} \end{aligned} \quad (10)$$

Since the parents of the variables in $\mathbf{Y}_{NC}$ do not include any $\mathbf{X}_i$, the conditional probability terms for the $\mathbf{Y}_{NC}$ are taken outside the integral in Equation (10). It follows that

$$p\left[y_i \mid pa'(Y_i)\right] = p\left[y_i \mid pa(Y_i)\right], \quad Y_i \in \mathbf{Y}_{NC} \quad (11)$$

and the remaining conditional probability terms must correspond to the integral in the last line of Equation (10):

$$\prod_{Y_i \in \mathbf{Y}_C} p\left[y_i \mid pa'(Y_i)\right] = \int_{\mathbf{X}_1} \ldots \int_{\mathbf{X}_{n_\mathbf{X}}} \prod_{Y_i \in \mathbf{Y}_C} p\left[y_i \mid pa(Y_i)\right] \prod_{X_i \in \mathbf{X}} f\left[\mathbf{x}_i \mid pa(\mathbf{X}_i)\right] d\mathbf{x}_1 \ldots d\mathbf{x}_{n_\mathbf{X}} \quad (12)$$

The right-hand side of Equation (12) corresponds to the general formulation in Equation (3). Therefore, we can use a SRM to solve for the joint probability of the $\mathbf{Y}_C$. In the case where all variables $\mathbf{Y}_C$ are defined as domains in the space of their continuous parents, we can reformulate Equation (12) to read

$$\begin{aligned} \prod_{Y_i \in \mathbf{Y}_C} p\left[y_i^{(k_i)} \mid pa'(Y_i)\right] &= \int_{\mathbf{x} \in \Omega(\mathbf{x})} f(\mathbf{x} \mid \mathbf{y}_P) d\mathbf{x} \\ \Omega(\mathbf{x}) &= \bigcap_{Y_i \in \mathbf{Y}_C} \Omega_{i,k_{pa}}^{(k_i)}(\mathbf{x}) \end{aligned} \quad (13)$$

where $f(\mathbf{x}|\mathbf{y}_P) = \Pi_{\mathbf{X}_i \in \mathbf{X}} f[\mathbf{x}_i | pa(\mathbf{X}_i)]$, with $\mathbf{Y}_P$ denoting the set of discrete parents to any of the continuous variables, $\mathbf{Y}_P = \{\mathbf{Y} \cap [pa(\mathbf{X}_1) \cup \ldots \cup pa(\mathbf{X}_{n_\mathbf{x}})]\}$. Note that Equation (13) has the form of Equation (1). If the probabilities in the rBN are computed directly using this equation, then the problem must be solved for all combinations of the states of the variables $\mathbf{Y}_C$, $\mathbf{Y}_P$ and $pa(\mathbf{Y}_C)$. This number can become prohibitively large in general eBN models. However, under certain circumstances, it is possible to take advantage of the independence assumptions encoded in the rBN to limit the number of SRM calculations. This is the case if it is possible to reformulate the integral in Equation (13) into the form

$$\int_{\mathbf{x} \in \Omega(\mathbf{x})} f(\mathbf{x}|\mathbf{y}_P) d\mathbf{x} = \prod_l \int_{\mathbf{x}_l \in \Omega_l(\mathbf{x}_l)} f(\mathbf{x}_l|\mathbf{y}_{Pl}) d\mathbf{x}_l$$

$$\Omega_l(\mathbf{x}_l) = \bigcap_{Y_i \in \mathbf{Y}_{Cl}} \Omega_{i,k_{pa}}^{(k_i)}(\mathbf{x}_l)$$

(14)

In Equation (14), the continuous random variables are separated into groups $\mathbf{X}_l$, with corresponding discrete children $\mathbf{Y}_{Cl}$ and discrete parents $\mathbf{Y}_{Pl}$, such that the SRM calculations can be performed separately for these groups. The total number of SRM calculations is then given by the product of the number of states of $\mathbf{Y}_{Cl}$, $\mathbf{Y}_{Pl}$ and $pa(\mathbf{Y}_{Cl})$, summed over all groups $l$. In the following section, we show how groups can be identified for which the decomposition in Equation (14) holds.

## 3.4 The Markov envelope

In the preceding section, a method was presented for reducing the eBN to a rBN by means of a SRM. It was postulated that the number of SRM calculations can be reduced by identifying groups of continuous random variables $\mathbf{X}_l$ for which such calculations can be performed separately. In this section we show that these groups are uniquely defined by the graphical structure of the eBN, and that the minimum number of SRM calculations required, therefore, follows directly from the graphical structure of the eBN and the number of states of its discrete variables.

First, we observe that if a continuous node $\mathbf{X}_j$ has no other continuous node in its Markov blanket, $bl(\mathbf{X}_j)$ as defined in Equation (6), then this node can be treated separately in the process of establishing the rBN. We can show this by considering the general formulation for the rBN, Equation (10). The rBN was obtained by integrating out the continuous variables. If a continuous node $\mathbf{X}_j$ has no other continuous node in its Markov blanket, then $pa(\mathbf{X}_j)$ as well as $sp(Y_j)$ do not contain any other continuous node. Furthermore, no continuous node can have $\mathbf{X}_j$ as its parent. For these reasons, we can separate the integration over $\mathbf{X}_j$ from the integration over the other continuous variables in (10). We can also separate the corresponding terms of the rBN on the left-hand side of (12) and write a separate equation for all terms involving $\mathbf{X}_j$:

$$\prod_{Y_i \in ch(\mathbf{X}_j)} p\left[y_i \middle| pa'(Y_i)\right] = \int_{\mathbf{X}_j} \prod_{Y_i \in ch(\mathbf{X}_j)} p\left[y_i \middle| pa(Y_i)\right] f\left[\mathbf{x}_j \middle| pa(\mathbf{X}_j)\right] d\mathbf{x}_j \qquad (15)$$

Since $pa'(Y_i)$ can only include variables that appear on the right-hand side of (15), all variables in $pa'(Y_i)$ are part of the Markov blanket of $\mathbf{X}_j$. This proves that we can treat $\mathbf{X}_j$ together with $bl(\mathbf{X}_j)$ as a separate eBN when establishing the rBN. It also follows that it is immaterial if the links from $\mathbf{X}_j$ are reversed before or after the links from other continuous variables, which are outside the Markov blanket of $\mathbf{X}_j$.

Second, it follows from the node elimination algorithm that: a) all discrete nodes in $ch(\mathbf{X}_i)$ will have all parents of $\mathbf{X}_i$ as parents in the rBN; b) the discrete node in $ch(\mathbf{X}_i)$ whose link from $\mathbf{X}_i$ is reversed last will have all other discrete variables in $ch(\mathbf{X}_i)$ as well as all discrete variables in $sp(\mathbf{X}_i)$ as parents, i.e., the node will have as parents all other discrete variables in $bl(\mathbf{X}_i)$; c) if $bl(\mathbf{X}_i)$ includes other continuous variables, then, after removal of $\mathbf{X}_i$, these will have all variables of $bl(\mathbf{X}_i)$ in their respective Markov blankets. As a consequence, after elimination of a second variable $\mathbf{X}_j$, one node in $ch(\mathbf{X}_j)$ will have all other nodes that were part of the Markov blankets of $\mathbf{X}_i$ and $\mathbf{X}_j$ in the eBN as parents. More generally, consider a set of continuous nodes $\mathbf{X}_M$, which is identified as follows: Start with a single continuous node and put it into $\mathbf{X}_M$; add all continuous nodes that are part of the Markov blanket of the first node; add all continuous nodes that are part of the Markov blankets of the additional nodes; and so on. We then define as a *Markov envelope* the aggregation of all variables (discrete and continuous) that are part of the Markov blankets of all variables in $\mathbf{X}_M$, i.e., $\{\bigcup_{\mathbf{X}_i \in \mathbf{X}_M} bl(\mathbf{X}_i)\}$. This concept is illustrated in Figure 3. It follows from the above considerations that one discrete variable in each of the Markov envelopes will have all other discrete variables in the envelope as parents in the rBN. Furthermore, the continuous random variables within a Markov envelope form the minimum groups $\mathbf{X}_I$ for which the equality in Equation (14) holds.

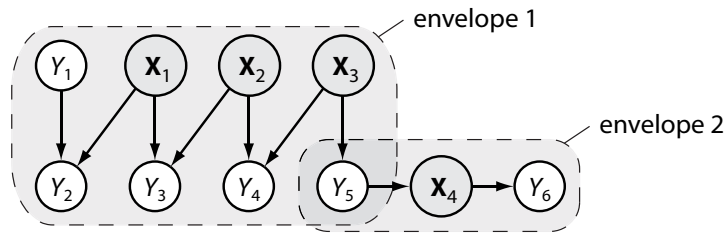

*Figure 3. Illustration of the principle of envelopes of Markov blankets of continuous variables.* $\mathbf{X}_1$ *and* $\mathbf{X}_3$ *are both in* $bl(\mathbf{X}_2)$, *thus the envelope contains* $bl(\mathbf{X}_1)$, $bl(\mathbf{X}_2)$ *and* $bl(\mathbf{X}_3)$, *whereas* $bl(\mathbf{X}_4)$ *contains no other continuous variables and forms an individual envelope.*

The fact that for one node $Y_i$ in each Markov envelope $pa'(Y_i)$ will include all other discrete variables in the envelope has fundamental implications for the resulting rBN. Independent of the ordering of link reversals, the sizes of these envelopes determine the number of SRM computations, since the potential of one node in each envelope will include all other discrete variables in the envelope. (The number of entries in the table is $\Pi_{i=1}^{n} m_i$, with $n$ being the number of discrete variables in the envelope and $m_i$ being the number of states of the $i$-th variable.) Furthermore, the maximum size of these envelopes represents a lower limit to the maximum clique size in the rBN. The importance of this observation stems from the fact that the maximum clique size is a crucial parameter for the computational speed in performing exact inference in BNs (Dechter, 1996).

It follows from the above analysis that, to ensure computational feasibility of the rBN, the number of discrete variables in any Markov envelope in the eBN must be limited. The maximum feasible number depends on $m_i$, but even in the extreme case of $m_i = 2$ for all $i$, the Markov envelope should not contain more than 15-20 discrete variables. In a later section, we discuss modeling strategies to deal with this problem.

## 3.5 Illustration

The derivation of the rBN is illustrated on the example depicted in Figure 2. Since this eBN contains only one node with continuous variables, there exists only one Markov envelope, consisting of the variables $\mathbf{X}_1$ and $Y_3 - Y_6$.

First, we demonstrate how the rBN shown on the far right of Figure 2 can be derived from the eBN through algebraic manipulations. The joint probability measure for this eBN is written as

$$p(y_1,\ldots,y_7|\mathbf{x}_1)f(\mathbf{x}_1) = \\ p(y_1)p(y_2|y_1)p(y_3|y_1,y_2)p(y_4|y_3)f(\mathbf{x}_1|y_3)p(y_5|y_4,\mathbf{x}_1)p(y_6|y_5,\mathbf{x}_1)p(y_7|y_5) \quad (16)$$

The rBN is obtained through integration over the domain of $\mathbf{X}_1$:

$$p(y_1,\ldots,y_7) = \int_{\mathbf{X}_1} p(y_1,\ldots,y_7|\mathbf{x}_1)f(\mathbf{x}_1)d\mathbf{x}_1 \\ = p(y_1)p(y_2|y_1)p(y_3|y_1,y_2)p(y_4|y_3)p(y_7|y_5)\int_{\mathbf{X}_1} f(\mathbf{x}_1|y_3)p(y_5|y_4,\mathbf{x}_1)p(y_6|y_5,\mathbf{x}_1)d\mathbf{x}_1 \quad (17)$$

with

$$\int_{\mathbf{X}_1} f(\mathbf{x}_1|y_3)p(y_5|y_4,\mathbf{x}_1)p(y_6|y_5,\mathbf{x}_1)d\mathbf{x}_1 = \int_{\mathbf{X}_1} p(y_5,y_6|y_4,\mathbf{x}_1)f(\mathbf{x}_1|y_3)d\mathbf{x}_1 \\ = p(y_5,y_6|y_4,y_3) \quad (18)$$

and inserting in (17) we obtain

$$p(y_1,...,y_7) = p(y_1) p(y_2|y_1) p(y_3|y_1,y_2) p(y_4|y_3) p(y_5,y_6|y_3,y_4) p(y_7|y_5) \qquad (19)$$

This formulation corresponds to the rBN obtained in Figure 2 by the node elimination algorithm, since $p(y_5,y_6|y_3,y_4) = p(y_5|y_6,y_3,y_4)p(y_6|y_3,y_4)$. The conditional probability tables of variables $Y_5$ and $Y_6$ in the resulting network must be computed. In accordance with (13),

$$p\left(y_5^{(k_5)}, y_6^{(k_6)} \Big| y_3^{(k_3)}, y_4^{(k_4)}\right) = \int_{\mathbf{x}_1 \in \{\Omega_{Y_5,k_4}^{(k_5)}(\mathbf{x}_1) \cap \Omega_{Y_6,k_5}^{(k_6)}(\mathbf{x}_1)\}} f\left(\mathbf{x}_1 \Big| y_3^{(k_3)}\right) d\mathbf{x}_1 \qquad (20)$$

Here, $\Omega_{Y_5,k_4}^{(k_5)}(\mathbf{x}_1)$ is the domain that defines the event $Y_5 = k_5$ in the space of $\mathbf{X}_1$ given that $Y_4 = k_4$, and $\Omega_{Y_6,k_5}^{(k_6)}(\mathbf{x}_1)$ is defined accordingly. Equation (20) can be solved using a system SRM. The individual potentials for $Y_5$ and $Y_6$ are then obtained simply by

$$p(y_5|y_3,y_4) = \sum_{Y_6} p(y_5,y_6|y_3,y_4) \qquad (21)$$

$$p(y_6|y_3,y_4,y_5) = \frac{p(y_5,y_6|y_3,y_4)}{p(y_5|y_3,y_4)} \qquad (22)$$

The derivation of the rBN for this example, therefore, requires solving $m_3 \cdot m_4 \cdot m_5 \cdot (m_6 - 1)$ system structural reliability problems (with $m_i$ being the number of states of $Y_i$).

### 3.6 Obtaining an optimal rBN from a given eBN

Although the minimum number of required SRM calculations to produce the rBN is determined by the structure of the eBN, it is possible to obtain different rBNs for a given eBN depending on the selected order of link reversals and elimination of continuous nodes. This is demonstrated by the example in Figure 4. In this section we briefly discuss the optimal ordering of the link reversal and node elimination actions.

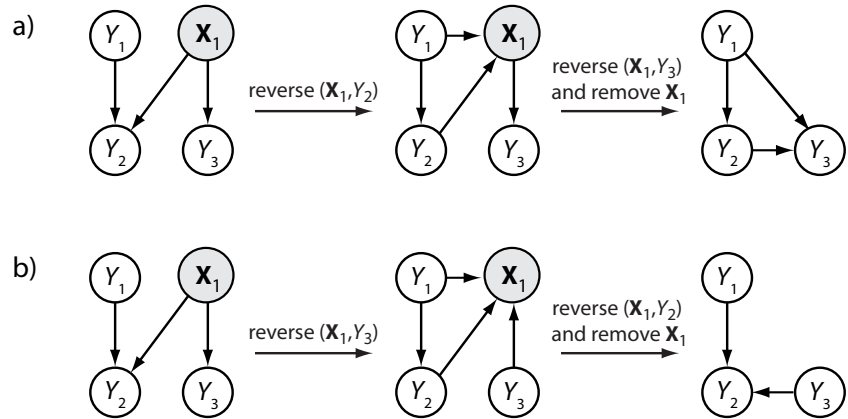

*Figure 4. The influence of the order of link reversals on the final rBN: Ordering (a) leads to an additional link in the rBN.*

The criterion for the optimality of the rBN depends on the envisioned application. Commonly, either the rBN leading to the lowest required CPU time or the one leading to the minimum storage requirement is considered as optimal. In other instances, computational or storage issues may not be as important as having a rBN that has links with logical (causal) interpretation. In most cases, however, these criteria will coincide. For example, in Figure 4, the ordering (b) leads to a rBN that is optimal according to all the above criteria.

It has been shown in a preceding section that each Markov envelope can be considered individually in the elimination algorithm. Therefore, only the orderings of link reversals and node eliminations *within* the Markov envelopes are relevant for the optimality of the rBN. Since the number of discrete variables within a Markov envelope must necessarily be limited, the number of combinations of link reversal orders to consider will generally be relatively small. In that case, the analyst may be able to determine the optimal rBN through inspection of the eBN graph. Based on observations made earlier, one can state that the link reversal should generally start with the links going to nodes with the fewest parents, if the goal is to have the rBN with a minimal number of links. Exceptions to this rule, however, may occur.

It is noted that the modeling choices made in establishing the eBN are more decisive for computational performance than the order of link reversals in establishing the rBN. This is because the form of the eBN determines the number of SRM calculations and presents a lower bound on the maximum clique size of the rBN. For this reason, emphasis should be placed on establishing a computationally efficient eBN model by focusing on the sizes of its Markov envelopes. This is discussed in a later section on modeling strategies.

## 3.7 Evidence and inference on continuous variables

A major motivation for the use of a BN is its capability for Bayesian updating when evidence, such as measurement results, monitoring data or observations of performances of structures, becomes available. Inclusion of evidence on any set of variables in the rBN is supported by the available exact inference algorithms for discrete-variable BNs. Therefore, in the eBN approach, potential evidence on continuous variables should somehow be represented in terms of discrete variables so they remain present in the rBN. To this end, when establishing the eBN, it is necessary to anticipate the type of evidences that may become available on continuous variables and introduce corresponding discrete variables.

Potential evidence for a group of continuous variables $\mathbf{X}_e$ may be described by a set of domains $\Omega_{e,i}(\mathbf{x}_e)$, $i = 1,...,m_e$, in which the variables $\mathbf{X}_e$ might be observed to fall. Let $p_i = \Pr[\mathbf{X}_e \in \Omega_{e,i}(\mathbf{x}_e)]$. Then a discrete variable $Y_e$ is introduced as a child of $\mathbf{X}_e$ with $m_e$ states that are defined through the domains $\Omega_{e,i}(\mathbf{x}_e)$. The evidence $\{\mathbf{X}_e \in \Omega_{e,i}(\mathbf{x}_e)\}$ is thus represented in the rBN as evidence $\{Y_e = i\}$ on $Y_e$. Note that the probabilities $p_i$ are of the form in Equation (1) and are easily computed by an SRM.

It is possible to envision evidence of zero probability. An event of the form $\{h(\mathbf{X}_e) = 0\}$, where $h$ denotes a deterministic function, has zero probability when $\mathbf{X}_e$ are continuous variables. In this case all SRM computations that involve $\mathbf{X}_e$ must be performed conditional on the zero-probability event. SRM enables such computations through surface integration (Schall *et al.*, 1988) or by reliability sensitivity analysis (Madsen, 1987). However, it might often be easier and more practical to represent such zero-probability observations with domains of small probability. For example, for the event mentioned above, one may consider the domain $\Omega_e(\mathbf{x}_e) = \{0 \leq h(\mathbf{x}_e) \leq \delta h\}$ with $\delta h$ a sufficiently small value. If observations of such a function over an interval are anticipated, then the interval needs to be discretized and corresponding domains introduced so that the method described in the preceding paragraph applies.

Since inference (updating) can be made only on variables that are present in the rBN, the outcome space of continuous variable on which inference is desired must be discretized. Methods for such discretization are described next.

## 3.8 Discretization of continuous random variables

Discretization of random variables in the eBN (or more precisely, discretization of the outcome spaces of continuous random variables) may be necessary for two reasons. First, if we are interested in the posterior distribution of a continuous random variable, its outcome space should be discretized to allow inference on the variable in the rBN. Second, an efficient strategy to reduce the size of the Markov envelopes is to selectively discretize continuous random variables, as we will demonstrate in the next section.

Surprisingly, little literature is available on discretization of continuous random variables in the context of engineering risk analysis, given that the approach is commonly used in engineering practice. Some considerations can be found in (Friis-Hansen, 2000; Neil *et al.,* 2008; Straub, 2009). There is need for a formalized approach, which, however, is beyond the scope of this paper. We limit ourselves to proposing an approach that is directly based on the eBN and which is accurate as well as practical for many applications.

Consider discretization of the outcome space of a continuous random variable $X_i$ with conditional cumulative distribution function $F_{X_i}[x_i | pa(X_i)]$. In the eBN, we replace $X_i$ by two random variables, a discrete variable $Y_i$ and a continuous variable $X_i'$, which is a child of $Y_i$. $Y_i$ inherits all parent variables of $X_i$, while $X_i'$ becomes the parent to all the children of $X_i$. This is illustrated in Figure 5. The outcome space of $Y_i$ consists of $m_i$ states, denoted by $y_i^{(k)}$, $k = 1\ldots m_i$. These correspond to mutually exclusive, collectively exhaustive intervals in the outcome space of the original variable $X_i$. The outcome space of $X'_i$ is identical to the outcome space of $X_i$.

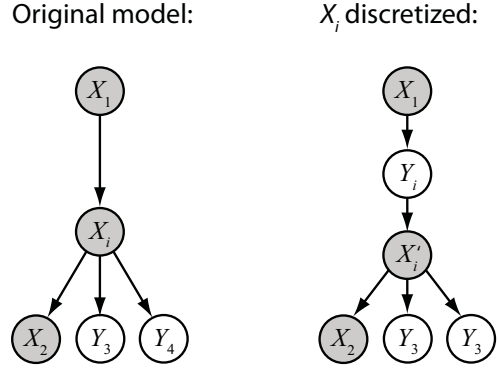

*Figure 5. Discretization of a random variable $X_i$ by replacing it with $Y_i$ and $X'_i$.*

The proposed approach to discretization makes direct use of the eBN framework. By maintaining a continuous random variable $X'_i$ in the eBN, it is not necessary to redefine the conditional distributions of the children of $X_i$; it suffices to replace the conditioning variable $X_i$ with $X'_i$. We must, however, determine the conditional PMF of $Y_i$ given its parents (the parents of $X_i$) and the distribution of $X'_i$ conditioned on $Y_i$. The continuous variable $X'_i$ is later eliminated in the process of establishing the rBN.

The conditional PMF of $Y_i$ is obtained as

$$p\left[y_i^{(k)}|pa(X_i)\right] = F_{X_i}\left[x_{ik}^+|pa(X_i)\right] - F_{X_i}\left[x_{ik}^-|pa(X_i)\right] \tag{23}$$

in which $x_{ik}^-$ and $x_{ik}^+$ are the lower and upper boundaries of the interval corresponding to state $k$ of $Y_i$. If $X_i$ is a deterministic function of its parents, $X_i = h_i[pa(X_i)]$, then the states of $Y_i$ can be defined directly as domains in the space of $pa(X_i)$: $\Omega_i^{(k_i)}[pa(x_i)] = \{x_{ik}^- - h_i[pa(x_i)] < 0\} \cap \{h_i[pa(x_i)] - x_{ik}^+ \leq 0\}$.

If $X_i$ has no parent, then the conditional distribution of $X'_i$ given $Y_i = y_i^{(k)}$ is obtained as

$$F_{X'_i}\left(x_i \mid y_i^{(k)}\right) = \begin{cases} 0, & x_i \leq x_{ik}^- \\ \dfrac{F_{X_i}(x_i) - F_{X_i}(x_{ik}^-)}{F_{X_i}(x_{ik}^+) - F_{X_i}(x_{ik}^-)}, & x_{ik}^- < x_i \leq x_{ik}^+ \\ 1, & x_{ik}^+ < x_i \end{cases} \qquad (24)$$

In this case, the discretization does not introduce any approximation, since the marginal distribution of $X'_i$ is identical to that of $X_i$. However, if $X_i$ has parents, then the marginal distribution of $X_i$ is not generally known and an approximation is required. A straightforward choice is the uniform distribution within each discretized interval $x_{ik}^-$ - $x_{ik}^+$, in which case

$$F_{X'_i}\left(x_i \mid y_i^{(k)}\right) = \begin{cases} 0, & x_i \leq x_{ik}^- \\ \dfrac{x_i - x_{ik}^-}{x_{ik}^+ - x_{ik}^-}, & x_{ik}^- < x_i \leq x_{ik}^+ \\ 1, & x_{ik}^+ < x_i \end{cases} \qquad (25)$$

The uniform assumption is not suitable if the discretization interval is bounded only on one side, as it occurs for the intervals in the extreme tails of unbounded distributions. In such cases, a different distribution must be selected. For example, for the unbounded interval $(x_{ik}^-, +\infty)$ one may select the exponential distribution (Straub, 2009)

$$F_{X'_i}\left(x_i \mid y_i^{(k)}\right) = \begin{cases} 0, & x_i \leq x_{ik}^- \\ 1 - \exp\left[-\lambda\left(x_i - x_{ik}^-\right)\right], & x_{ik}^- < x_i \end{cases} \qquad (26)$$

where $\lambda$ must be selected by the analyst to reflect the anticipated rate of decay of the tail of the marginal distribution.

Even when $X_i$ has parents in the eBN, its marginal distribution might be known and $F_{X'_i}(x_i \mid y_i^{(k)})$ can be determined from (24). In such cases, the discretization still entails an approximation, since $X'_i$ cannot fully reflect the distribution of $X_i$ within one interval of $Y_i$ for given values of $pa(X_i)$. In other words, even though the marginal distribution of $X_i$ is correctly represented, the statistical dependence between $X_i$ and its parents is only approximately represented in the discretization.

With the presented approach, all discrete children of the original continuous variable $X_i$ become children of $X_i'$, and, therefore, are part of the same Markov envelope. It can be desirable to have them in separate Markov envelopes, in order to reduce the number of SRM computations and to limit the complexity of the resulting rBN. This can be achieved by introducing a separate continuous random variable $X_{ij}'$ for each child of $X_i$, as illustrated in Figure 6. Here, all $X_{ij}'$ are defined through identical distributions conditional on $Y_i$ as given in Eqs. (24) to (26). This approach to discretization introduces an additional approximation in the model: In reality, the $X_{ij}'$ should be identical, but this model only considers that they are in the same interval as specified by $Y_i$. This approximation does not affect the model of the marginal distributions, but it leads to an underestimation of the statistical dependence among the children of $Y_i$.

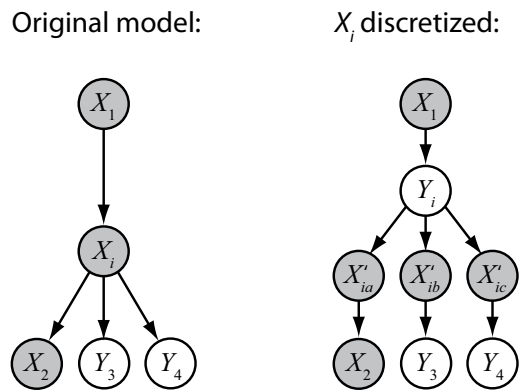

*Figure 6. Alternative discretization of a random variable $X_i$, separating the children of $X_i$.*

## 4 Modeling strategies

The proposed solution strategy for the eBN has two computational bottlenecks: The number of SRM computations necessary to determine the conditional probability tables, and the size of the largest clique in the resulting rBN. The number of SRM computations increases exponentially with the number of discrete variables in each Markov envelope. The maximum clique size has as a minimum value the number of discrete variables in each Markov envelope, but it also depends on the dependence structure of the discrete variables. As a consequence, to ensure computability of the rBN, the sizes of the Markov envelopes in the eBN must be limited. In this section, we discuss strategies for doing this.

*a) Discretization of continuous random variables*

Markov envelopes are separated when continuous variables are not directly linked and if they are not connected through common children, as illustrated in Figure 3. Therefore, an efficient strategy to reduce the size of Markov envelopes, and thus the number of SRM calculations and the complexity of the rBN, is to selectively discretize continuous random variables. This strategy is particularly effective in hierarchical eBN structures, as shown in Figure 7. If each $Y_i$, $i = 1,...,5$, has $m$ discrete states, then the evaluation of the rBN for the original model (a) requires $m^4(m-1)$ system SRM calculations, whereas model (b), wherein the continuous variable $X_0$ has been replaced with the discrete variable $Y_0$ and corresponding continuous variables $X'_{0a},..., X'_{0e}$, requires only $5(m-1)$ component SRM calculations. If the components are identically defined (e.g., five structural components of similar type), then the number of component SRM calculations is further reduced to $(m-1)$.

An additional example of how discretization reduces the size of Markov envelopes is demonstrated in Figure 8 for a dynamic eBN. Here, discretization of the variables connecting the different slices (corresponding to different instances of time or space) leads to a much simpler rBN structure. To this end, the continuous random variable $X_i$ is replaced by the discrete $Y_{ia}$ and corresponding continuous $X'_{ia}$ and $X'_{ib}$. The reduction in the number of required SRM calculations is identical to that of the example in Figure 7.

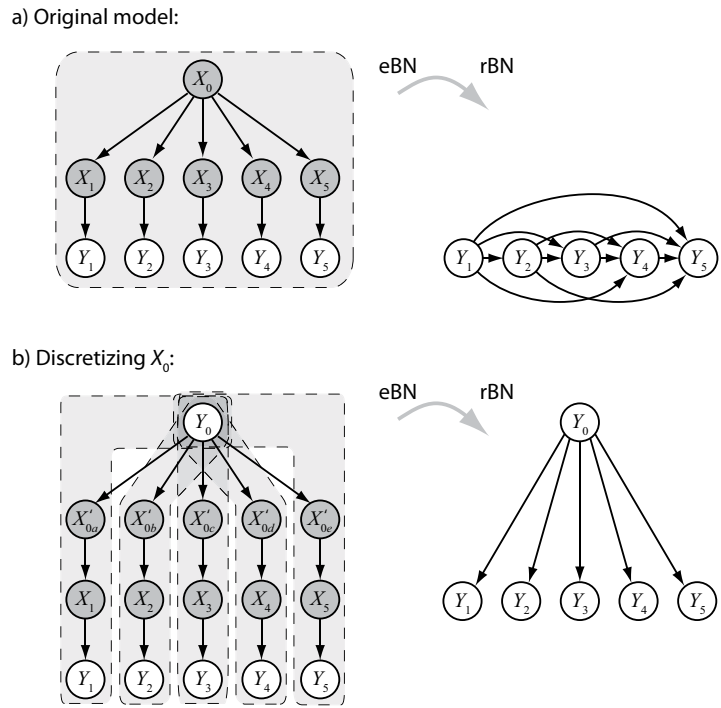

*Figure 7. Discretizing the common parent variable reduces the size of the Markov envelopes.*

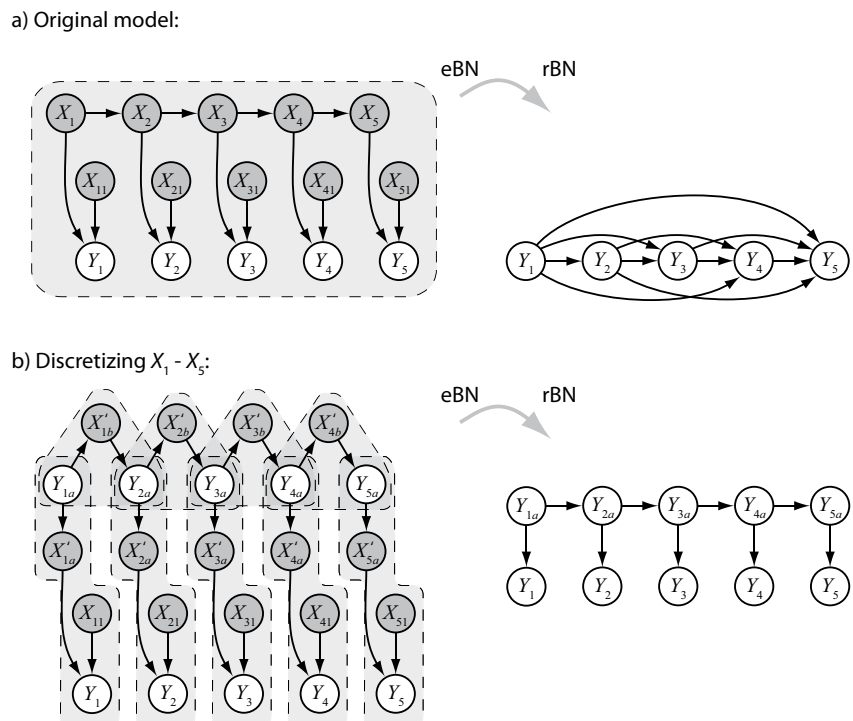

*Figure 8. Discretizing the interconnecting variables to reduce the size of the Markov envelopes in the dynamic eBN.*

*b) Causal and explicit modeling*

Because the rules for encoding the dependence structure in the BN graph are derived from causal reasoning (Pearl, 1988), causal modeling of the relations among variables generally leads to the lowest number of links in a BN. Figure 9 shows the classical example of two test outcomes $Y_1, Y_2$ of a system. If the test outcomes are modeled conditionally on the state of the system $Y_0$, as in the causal model on the left, then it can reasonably be assumed that $Y_1$ is statistically independent of $Y_2$ given $Y_0$. On the other hand, if the PMF of the system state is defined conditional on the test outcomes (the so-called diagnostic model on the right), then $Y_1$ and $Y_2$ become statistically dependent. (This becomes evident from applying the link reversal algorithm to the causal network.)

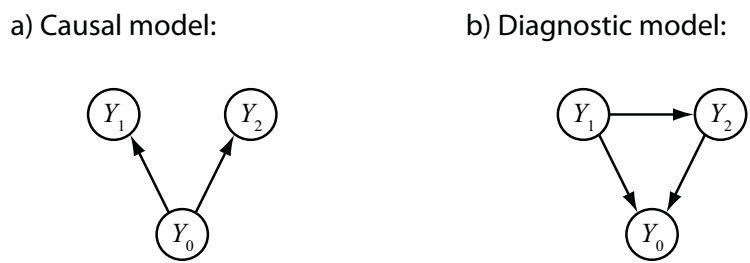

*Figure 9. Causal versus diagnostic modeling in the eBN. $Y_0$ is system state and $Y_1$ and $Y_2$ are two outcomes of tests on $Y_0$.*

Additionally, the complexity of the eBN and the resulting rBN can often be reduced by including variables explicitly as separate nodes in the eBN. Consider the example of a set of 5 equi-correlated random variables. In a direct representation of these variables the last variable would have all other variables as parents (similar to the rBN in Figure 7a). However, equi-correlation is typically caused by a common influencing factor. If such a factor is explicitly included in the model as a common parent node, then the dependence among the variables can be fully represented by a simple network structure, similar to the rBN in Figure 7b. This structure can represent equi-correlation among variables $Y_1$-$Y_5$ through the common factor $Y_0$.

*c) Maintaining causality in the rBN*

To reduce the complexity of the rBN and to simplify the SRM calculations, it is often beneficial to maintain causality in the rBN, even though this might require discretizing additional variables. An example is shown in Figure 10a, in which $X_1$ - $X_5$ represent random variables influencing the system/component performance $Y_0$. Discrete variables $Y_{1a}$ and $Y_{2a}$ are outcomes of tests performed on $X_1$ and $X_2$. If the continuous variables $X_1$ and $X_2$ are discretized, as shown in Figure 10b, the resulting rBN maintains causality. If the states of $Y_0$ are defined by single limit-state functions, it is then sufficient to calculate $p(y_0)$ through component SRM calculations to obtain the rBN. In the original model, Figure 10a, it is necessary to compute the joint PMF of $Y_0$, $Y_{1a}$ and $Y_{2a}$, which will necessitate system SRM calculations.

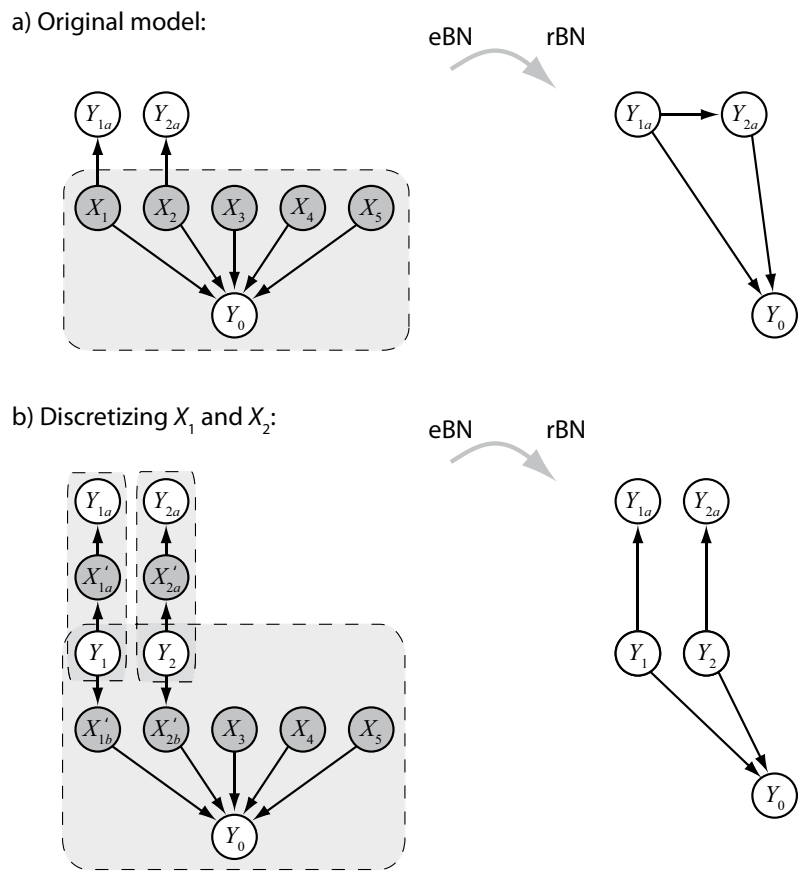

*Figure 10. By maintaining causality in the rBN, system SRM calculations might be avoided.*

*d) Divorcing of variables*

In some cases it is possible to reduce the number of parents to a variable by a divorcing strategy (Jensen and Nielsen, 2007). This strategy is useful when the joint influence of a number of parent variables can be represented by a single intermediate variable. As an example consider a case where the structural performance $Y_0$ is a function of four discrete variables $Y_1$-$Y_4$ and a number of continuous variables $\mathbf{X}$. Suppose the failure state is described by the limit-state function $g = y_4 - a_1 y_1 - a_2 y_2 - a_3 y_3 + h(\mathbf{x})$, where $h(\mathbf{x})$ is any function and $a_1$-$a_3$ are deterministic coefficients. In this case, the number of parents to $Y_0$ can be reduced to three by introducing the new variable $Y_5 = a_1 Y_1 + a_2 Y_2 + a_3 Y_3$, as illustrated in Figure 11.

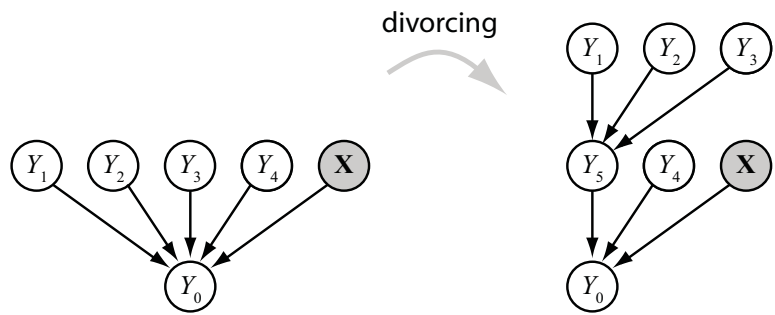

*Figure 11. Illustration of the divorcing strategy* (here, parents $Y_4$ and $\mathbf{X}$ are "divorced" from the remaining discrete parents of $Y_0$).

As a final remark, we note that the eBN approach exploits conditional independence among random variables in a probabilistic model. Therefore, the approach does not present advantages for modeling problems that do not exhibit such independence. Consider a discretized random field represented by a vector of dependent variables $\mathbf{X}$. Such a vector is not Markovian and cannot be represented by a hierarchical BN structure as in Figure 7b. If observations are available at $n$ locations modeled through $n$ discrete variables $Y_1$-$Y_n$, one variable in the resulting rBN will have all other $n-1$ variables as parents. Such a model can become computationally impractical even for a moderate number of observed variables. (Remember that the conditional probability table of the last node will have $m^n$ entries, with $m$ being the number of discrete states of each variable $Y_i$.) This fundamental inability to efficiently model complex dependence structures among discrete random variables that are not characterized by causal relations is the main limitation of the eBN approach. Although ideas such as using principal component analysis to reduce the dependence structure have been explored (Straub *et al.,* 2008), further work is required to address this problem within the eBN framework. Luckily, most models used in civil engineering involve causal relations among the variables.

## 5  Summary and Conclusions

This paper presents the idea of using structural reliability methods (SRMs) for solving enhanced Bayesian networks (eBNs), which include both continuous and discrete random variables whose states can be described by sets of limit state functions. The proposed approach employs a node elimination algorithm, which removes the continuous nodes to arrive at a reduced BN, the rBN, which only has discrete nodes and can be solved by existing exact inference algorithms. SRMs are used to compute the conditional probability tables of the rBN as component or system reliability problems. To evaluate the number of SRM calculations and the complexity of the resulting rBN the concept of Markov envelopes is introduced, and it is shown that the computational requirements are determined by the sizes of the Markov envelopes. A number of modeling strategies are described to reduce the sizes of the Markov envelopes, and thereby reduce the number of required SRM calculations and the maximum clique size of the rBN. These strategies include selectively discretizing continuous variables, maintaining causal relations between the nodes, and divorcing parent nodes. Additionally, the problems of entering evidence on continuous variables and inference on continuous variables are addressed.

An alternative for dealing with the eBN that has not been considered here is to perform approximate inference directly on the eBN, e.g., by means of MCMC. The potential advantage of such an approach lies in the ability to handle more general forms of dependence among the variables, including those arising from random fields. However, it is important to realize that approximate inference has its own limitations, in particular when the interest is in computing probabilities of rare events described by limit-state functions, and when the interest is in near-real-time inference and decision analysis.

The theoretical framework presented in this paper has many potential applications, especially for decision support in near-real-time under uncertain and evolving information. Example areas of such application include early warning systems, emergency response and recovery planning for natural hazards, and the optimization of inspection, monitoring and repair actions in infrastructure systems. The companion paper presents two such applications related to structural and infrastructure systems.

## References

Beck, J. L., and S.-K. Au (2002), Bayesian Updating of Structural Models and Reliability using Markov Chain Monte Carlo Simulation, *Journal of Engineering Mechanics, Trans. ASCE,* **128**(4), 380–391.

Bensi, M. T., D. Straub, P. Friis-Hansen, and A. Der Kiureghian (2009), Modeling infrastructure system performance using BN, in *Proc. ICOSSAR'09,* Osaka, Japan.

Bobbio, A., L. Portinale, M. Minichino, and E. Ciancamerla (2001), Improving the analysis of dependable systems by mapping fault trees into Bayesian networks, *Reliability Engineering & System Safety,* **71**(3), 249–260.

Cooper, G. F. (1990), The computational complexity of probabilistic inference using Bayesian belief networks, *Artificial Intelligence,* **42**(2-3), 393–405.

Dechter, R. (1996), Bucket elimination: A unifying framework for probabilistic inference, in *Twelthth Conference on Uncertainty in Artificial Intelligence,* Portland, Oregon.

Der Kiureghian, A. (2005), First- and second-order reliability methods. Chapter 14, in *Engineering design reliability handbook,* edited by E. Nikolaidis et al., CRC Press, Boca Raton, FL.

Ditlevsen, O., and H. O. Madsen (1996), *Structural Reliability Methods,* John Wiley & Sons.

Ellingwood, B. R. (2006), Structural safety special issue: General-purpose software for structural reliability analysis, *Structural Safety,* **28**(1-2), 1–2.

Faber, M. H., I. B. Kroon, E. Kragh, D. Bayly, and P. Decosemaeker (2002), Risk Assessment of Decommissioning Options Using Bayesian Networks, *Journal of Offshore Mechanics and Arctic Engineering,* **124**(4), 231–238.

Friis-Hansen, A. (2000), Bayesian Networks as a Decision Support Tool in Marine Applications, *PhD thesis,* DTU, Lyngby, Denmark.

Friis-Hansen, P. (2004), Structuring of complex systems using Bayesian network., in *Proceedings Workshop on Reliability Analysis of Complex Systems,* Technical University of Denmark, Lyngby.

Gilks, W. R., S. Richardson, and D. J. Spiegelhalter (1996), *Markov chain Monte Carlo in practice,* Chapman & Hall, London.

Grêt-Regamey, A., and D. Straub (2006), Spatially explicit avalanche risk assessment linking Bayesian networks to a GIS, *Natural Hazards and Earth System Sciences,* **6**(6), 911–926.

Jensen, F. V., and T. D. Nielsen (2007), *Bayesian Networks and Decision Graphs.* Information Science and Statistics, Springer, New York, NY.

Langseth, H., T. D. Nielsen, R. Rumí, and A. Salmerón (2009), Inference in hybrid Bayesian networks, *Reliability Engineering and System Safety,* **94**(10), 1499–1509.

Langseth, H., and L. Portinale (2007), Bayesian networks in reliability, *Reliability Engineering & System Safety,* **92**(1), 92–108.

Lauritzen, S. L., and D. J. Spiegelhalter (1988), Local Computations with Probabilities on Graphical Structures and Their Application to Expert Systems, *Journal of the Royal Statistical Society. Series B (Methodological),* **50**(2), 157–224.

Madsen, H. O. (1987), Model Updating in Reliability Theory, in *Proc. ICASP 5,* Vancouver, Canada.

Mahadevan, S., and R. Rebba (2005), Validation of reliability computational models using Bayes networks, *Reliability Engineering & System Safety,* **87**(2), 223–232.

Murphy, K. P. (2001), The Bayes Net Toolbox for Matlab, *Computing Science and Statistics,* **33**.

Murphy, K. P. (2002), Dynamic Bayesian networks: representation, inference and learning, *PhD thesis,* University of California, Berkeley.

Neil, M., M. Tailor, D. Marquez, N. Fenton, and P. Hearty (2008), Modelling dependable systems using hybrid Bayesian networks, *Reliability Engineering & System Safety,* **93**(7), 933–939.

Nishijima, K., M. A. Maes, J. Goyet, and M. H. Faber (2009), Constrained optimization of component reliabilities in complex systems, *Structural Safety,* **31**(2), 168–178.

Pearl, J. (1988), *Probabilistic reasoning in intelligent systems : networks of plausible inference.* The Morgan Kaufmann series in representation and reasoning, Morgan Kaufmann Publishers, San Mateo, Calif.

Rackwitz, R. (2001), Reliability analysis – a review and some perspectives, *Structural Safety,* **23**(4), 365–395.

Russell, S. J., and P. Norvig (2003), *Artificial intelligence: a modern approach,* Prentice-Hall, Englewood Cliffs, N.J.

Schall, G., S. Gollwitzer, and R. Rackwitz (1988), Integration of multinormal densities on surfaces, in *Proc. 2nd IFIP WG 7.5 Working Conference,* London.

Shachter, R. D. (1986), Evaluating Influence Diagrams, *Operations Research,* **34**(6), 871–882.

Shachter, R. D. (1988), Probabilistic Inference and Influence Diagrams, *Operations Research,* **36**(4), 589–604.

Shenoy, P. P., and G. Shafer (1990), Axioms for probability and belief-function propagation, in *Proceedings of the Fourth Annual Conference on Uncertainty in Artificial Intelligence,* North-Holland Publishing Co.

Straub, D. (2009), Stochastic Modeling of Deterioration Processes through Dynamic Bayesian Networks, *Journal of Engineering Mechanics, Trans. ASCE*, **135**(10), pp. 1089-1099.

Straub, D., M. T. Bensi, and A. Der Kiureghian (2008), Spatial Modeling of Earthquake Hazard and Infrastructure Performance Through Bayesian Networks, in *Proc. EM'08 conference,* University of Minnesota, Minneapolis.

Straub, D., and A. Der Kiureghian (2010), Combining Bayesian Networks with Structural Reliability Methods: Application, *Journal of Engineering Mechanics, Trans. ASCE*, **136**(10): 1259-1270.

Wen, Y. K., and H. C. Chen (1987), On fast integration for time variant structural reliability, *Probabilistic Engineering Mechanics,* **2**(3), 156–162.

Zhang, N. L., and D. Poole (1996), Exploiting Causal Independence in Bayesian Network Inference, *Journal of Artificial Intelligence Research,* **5,** 301–328.